# State-of-the-Art Flash Chips for Dosimetry Applications

Preeti Kumari, Levi Davies, Narayana P. Bhat, En Xia Zhang, *Senior Member, IEEE,* Michael W. McCurdy, *Senior Member, IEEE*, Daniel M. Fleetwood, *Fellow, IEEE,* and Biswajit Ray, *Member, IEEE*

*Abstract*—In this paper we show that state-of-the-art commercial off-the-shelf Flash memory chip technology (20 nm technology node with multi-level cells) is quite sensitive to ionizing radiation. We find that the fail-bit count in these Flash chips starts to increase monotonically with gamma or X-ray dose at 100 rad($SiO_2$). Significantly more fail bits are observed in X-ray irradiated devices, most likely due to dose enhancement effects due to high-Z back-end-of-line materials. These results show promise for dosimetry application.

## I. Introduction

AUTONOMOUS, real-time radiation sensing with high precision is an important topic for several applications including military, civilian health and safety, space-exploration, etc. There exist different types of radiation dosimeters, such as small finger/body film badges, thermo-luminescent dosimeters (TLDs) [1], optically stimulated luminescence dosimeters (OSLs) and electronic field effect transistor based dosimeters (RADFETs) [2]–[5], each with its own strengths and weaknesses [6]. Many types of dosimeters need to be sent to a laboratory to be evaluated for exposure level and hence cannot perform real-time in situ measurement of the radiation environment.

Flash memory chips are low cost, high density with small-footprint and widely used in many embedded systems such as smartphones, etc. Flash chips store information in the form of charge on its floating gate (see Fig. 1(a)). A floating gate metal-oxide-semiconductor field effect transistor (FG-MOSFET), allows electrons from the silicon substrate to tunnel into the floating gate during the program operation. Under ionizing radiation exposure, charge loss takes place from the floating gate of the Flash cell through the mechanisms [7]–[10] shown in Fig 1(b). Charge loss decreases the programmed cell threshold voltage ($V_t$) distribution (see Fig. 1(c)), resulting in "0"→ "1" fails [11]–[13].

The key idea behind Flash-based dosimetry is to correlate the radiation induced fail bit count (FBC) with the radiation dose [6]-[9]. In 1998 Scheick et al. [14] demonstrated the utility of Electrically Erasable Programmable Read-Only Memory (EEPROM) for measuring ionizing radiation in space through the Microelectronics and Photonics Test Bed (MPTB) satellite. More recently, Savage et al. [15] proposed extreme value analysis in order to use floating gate memory as a dosimeter. In this paper we use multiple Flash chips of 20 nm technology node from Micron Technology in order to demonstrate its usability as a dosimeter. Our key contributions in this paper are as follows:

1) We find that commercial un-modified MLC (multi-level cell) Flash chips can detect radiation doses as low as 100 rad($SiO_2$) during X-ray and gamma-ray exposure.
2) We propose and demonstrate different techniques in order to improve radiation sensitivity and minimize measurement errors.
3) We propose an algorithm for the development of a smart phone application on the Android platform based on the radiation response of Flash characteristics in order to demonstrate the possibility of continuous real-time monitoring of the radiation dose.

## II. Experimental Details

We have utilized commercial off the shelf NAND Flash memory chips of 20 nm technology node from Micron Technology (part # MT29F64G08CBABAWP: B TR) in TSOP (Tape and Reel) packages. The chips are of size 64Gb with MLC storage. In the experiments we have used 100 different blocks from different physical locations of the chip. Each block consists of 256 pages of size 8k bytes each. A custom-designed board is used in order to program and read the Flash chips. The board contains a socket to hold a Flash chip under test, an ARM microprocessor to issue commands and receive data from the Flash chip, and a serial interface.

X-ray irradiation was performed using an ARACOR Model 4100 10-keV X-ray irradiator at a dose rate of 5 krad($SiO_2$)/min. All the terminals of the device under test (DUT) were grounded during exposure. Gamma-ray experiments were performed using a Cs-137 isotopic irradiator at a dose rate of 20 rad($SiO_2$)/min. Both X-ray and Gamma-ray exposures [16] were performed in the radiation test facilities of Vanderbilt University.

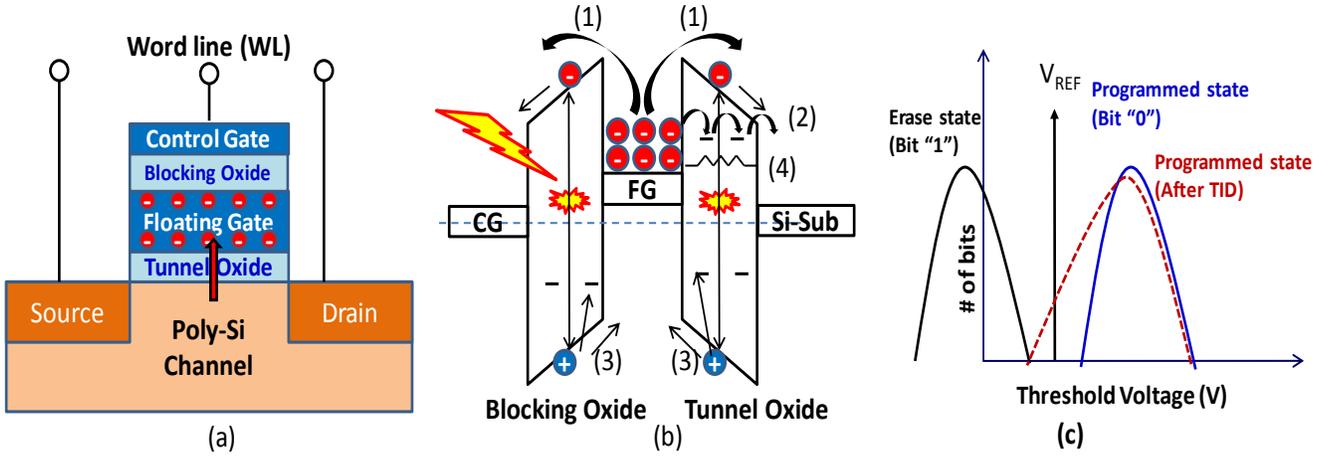

Fig. 1. (a) Schematic diagram of a floating gate Flash memory cell. (b) Energy band diagram with different possible pathways for charge loss after irradiation: (1) electron emission through the tunnel oxide and blocking oxide, (2) trap assisted tunneling, (3) generation of electron-hole pairs in blocking/tunnel oxide and subsequent recombination or hole trapping, and (4) conductive pipe model for charge loss. (c) Threshold voltage ($V_t$) distribution of the memory cells in a page/block. After ionizing radiation exposure, the program state $V_t$ shifts, causing fail bits. (After [7]-[13].)

## III. RESULTS AND ANALYSIS

In order to evaluate the effects of X-ray exposure on the Flash chips, we first remove the packaging material from the TSOP chip, as shown in Figs. 2(a) and (b). We then program the entire chip with an "ALL-0" data pattern. Just after the program, if we read the chip back, we find that there are a few bits already in the failed state ("0"→"1" bit flip shown as blue dots in Fig. 2(c)). These fail bits are inherent in the MLC (multi-level-cell) chips of advanced technology nodes due to very minimal voltage margin between the programmed states [17]. This inherent fail bit count (FBC) in a page of 8k bytes size is very small (< 10 bits per page), and can be easily corrected by the standard error correction (ECC) engine. The inherent FBC remains almost the same or slightly increases (over a few months of time) at room temperature. Once the chip is exposed to ionizing radiation (X-ray in the case of Fig. 2), the FBC increases significantly as a function of radiation dose, as shown in Fig. 2(c). Each data point in Fig. 2(c) is the fail bit count per page (page number shown on x-axis) after radiation exposure. We find that the radiation induced FBC increase varies from page to page within a block, as shown in the scatter plot in Fig. 2(c). This intra-block variation is due to the inherent differences between the pages within a block. However, the block average FBC remains approximately the same for all the blocks of the same chip (see Fig. 3(a)), which is irradiated uniformly. This implies that, for the purpose of radiation dosimetry, reading of a few blocks in a chip may provide sufficient accuracy.

In Fig. 3(b) we show the cumulative probability distribution function (CDF) for FBC/page in the chip as a function of X-ray dose. The steepness of the CDF curves indicates minimal variation in the measured block-averaged FBC. In Fig. 3(c) we plot the functional relationship between average FBC and the X-ray radiation dose. These results can be used for calibration purposes for the Flash based dosimetry for the X-rays. In the same plot we also compare the radiation response of the Flash chip with and without its package (or capping). The chip is more sensitive to X-ray irradiation without any capping on it. This is because the packaging material blocks most of the ~ 10-keV X-rays, allowing only the high energy X-rays in the tail of the distribution to penetrate inside the package [18].

We have also performed experiments in order to quantify the total ionization response using gamma rays (Cs-137 isotopic irradiator). With gamma rays we were able to control the expose the device in dose steps of 100 rad($SiO_2$). In Fig. 4 we plot the evaluation results for the effects of gamma-rays on the Flash chip of the same specification. Similar to the X-ray exposure, the FBC on the freshly written chip increases with the increase of gamma-ray dose.

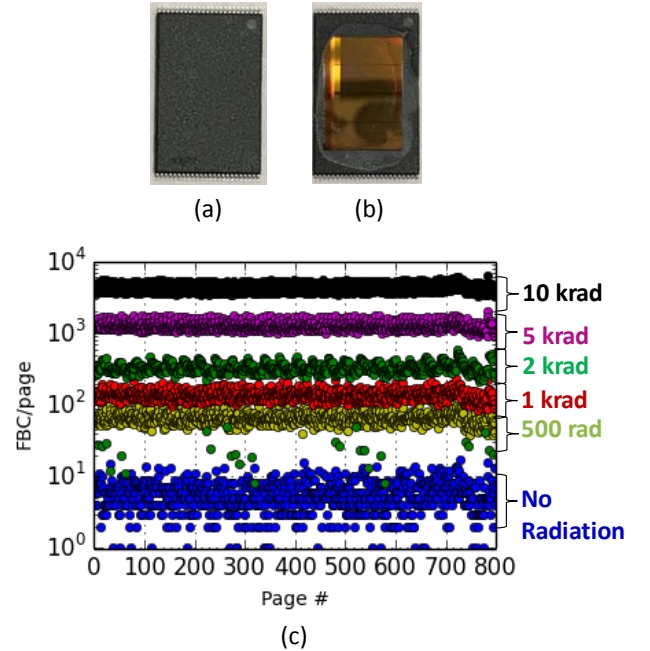

Fig. 2. (a) Flash memory chip with packaging on. (b) De-capped Flash chip. (c) Scatter plot of fail bits on different pages of the same chip as a function of radiation (X-ray) dose.

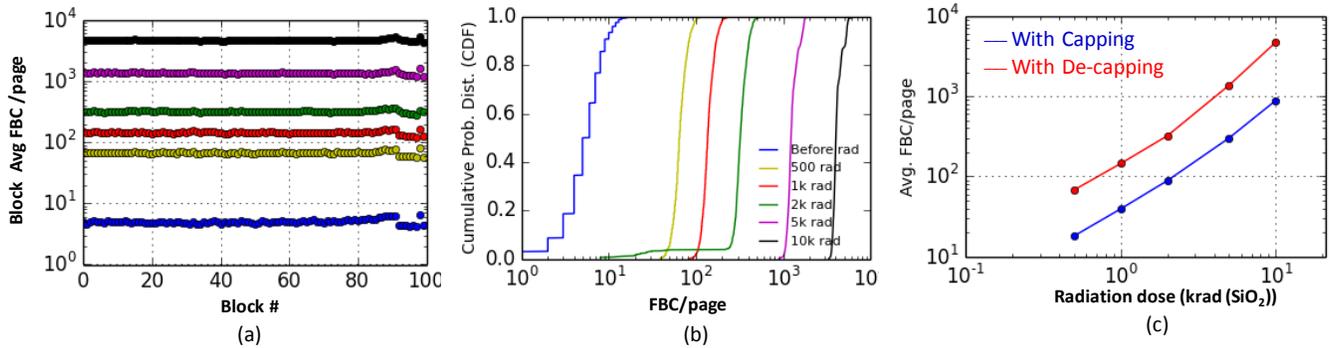

Fig. 3. Fail bit count for X-ray exposure of a Flash memory chip. (b) Block average FBC/page for each block is plotted as a function of radiation dose. The color coding for the dose levels is the same as in (a). (c) The cumulative probability distribution of the FBC for each radiation dose level is plotted. The legend in the plot shows the dose level. The FBC value just after programing is very similar in both chips.

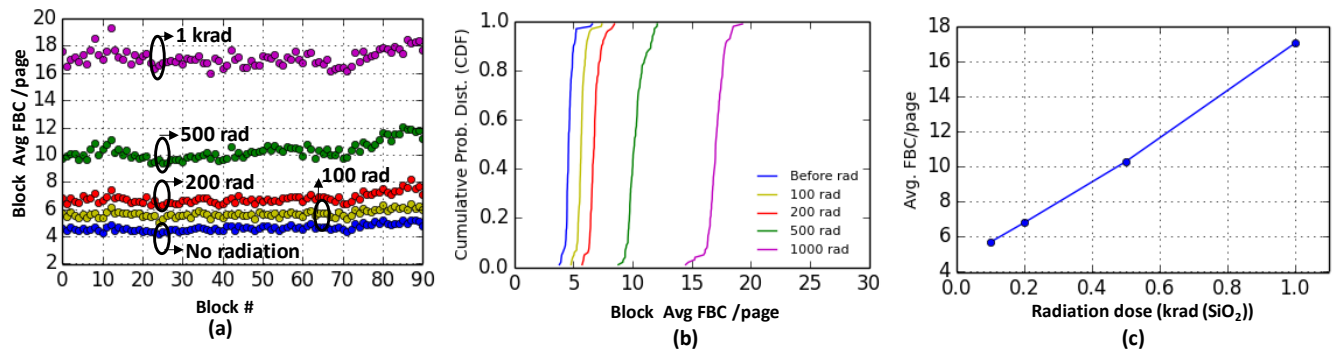

Fig. 4. Fail bit count for gamma-ray exposure of a Flash memory chip. (a) Increase in FBC with radiation dose. (b) The cumulative probability distribution of the FBC for each radiation dose level. (c) Average FBC in a chip as a function of radiation dose starting with 100 rad($SiO_2$).

Note in Fig. 4 that the MLC chip is sensitive enough to show radiation damage even with 100 rad($SiO_2$) of exposure, which can be measured with standard digital interfaces. These results imply that the state of the art Flash chip with lower technology nodes are quite sensitive to total-ionizing-dose effects. This limits their potential use in space environments, but is promising for low intensity radiation dosimetry.

The fail bit count in the X-ray exposure is significantly higher than in the gamma ray exposure for the same dose. For example, ~ 120 ± 10 FBCs are observed in Fig. 3 for X-ray irradiation to 1 krad($SiO_2$), while only 17 ± 1 FBCs are observed for gamma-ray irradiation to 1 krad($SiO_2$) in Fig. 4. The higher FBC during X-ray irradiation is due most likely to dose enhancement effects [16], [19]–[21] associated with the presence of high-Z materials (e.g., W) that are typically present in the back end of the line (BEOL) layers of state-of-the art memory chips [22]–[24]. Dose enhancement effects of up to 5-10 times can be observed for devices with similar high-Z BEOL materials irradiated with 10-keV X-rays [18],[25], consistent with these results. More work is required to understand and quantify these effects for different types of radiation sources.

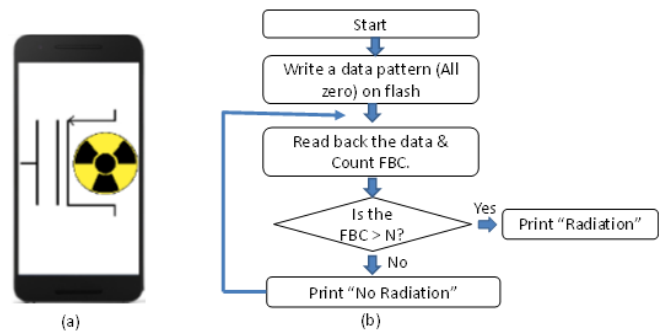

Fig 5. (a) A prototype for radiation dosimetry application. (b) Algorithm for reporting radiation alert by the application.

Since Flash memory is frequently used in modern smart-phone devices, the results in this paper encourage us to develop an application program for smart-phones that can autonomously record the radiation exposure. In Fig. 5 we propose an algorithm that can utilize radiation-induced memory errors for real-time dosimetry. The algorithm is implemented in an Android platform using Android Studio IDE (Fig. 5(a)), targeted towards a Samsung Galaxy S5 testbed chosen as a median representative of recent Android devices. The application is written in the Android language, similar to the more prevalent Java. SD cards of the phone is used for the flash memory. The basic form of the algorithm developed is modeled in the flowchart of Fig. 5(b). The Android operating system imposes limits on how low-level an application can read, and as a result it is difficult to acquire raw FBC data from the SD card. Instead, the application can periodically attempt to read a given file. If the file becomes corrupted, the current application can so far only indicate that a potentially high dose has been received. Flash memory is not sensitive enough for personnel dosimetry, but such an application could be used for inexpensive sensitive-area monitoring.

## IV. Conclusion

In this paper we demonstrate that the state of the art Flash chips are sensitive enough to detect low radiation doses such as 100 rad($SiO_2$). We find that X-ray exposure causes significantly a higher fail bit count compared to gamma ray exposure, mostly likely due to dose enhancement effects associated with high-Z BEOL materials. In general, we find that the radiation-induced fail bit count varies between different pages, but the block-averaged failure count remains consistent among all the blocks. Thus, for dosimetry applications, average failure count on any block correlates with dose. We have also developed an algorithm for smart phone application that can potentially be used for real-time radiation dosimetry. Finally, we note that the observed extreme radiation sensitivity of these devices would be quite limiting for potential space applications.